  \documentstyle[12pt]{article}
\begin{document}
\pagestyle{empty}
\parindent 0pt
\null\vskip-40pt
\vfill
\centerline{\bf First-order phase transition in $1d$ Potts model with long-range interactions}  
\vskip .25in
\centerline{K Uzelac and Z Glumac}
\vskip .25in
{\it Institute of Physics, Bijeni\v cka 46, POB 304,  10000 Zagreb, Croatia}  
\vskip 2.5 cm
\centerline{\bf Abstract}

\baselineskip 21pt

The first-order phase transition in the one-dimensional $q$-state Potts model with 
long-range interactions
decaying with distance as $1/r^{1+\sigma}$ 
has been studied
by Monte Carlo numerical simulations for $0 < \sigma \le 1$ and integer values of $q > 2$.
On the basis of finite-size scaling analysis of interface free energy $\Delta F_L$, specific
heat and Binder's fourth order cumulant, we obtain the first-order transition which occurs  
for $\sigma$ below a threshold value $\sigma_c(q)$.

\vskip 2cm

Physics Abstracts classification number: 05.50.+q, 64.60.Cn

\vfill
\eject
\pagestyle{plain}
\parindent 6mm

The subject of our study is the one-dimensional ($1d$) Potts model with ferromagnetic 
long-range (LR) interactions decaying with distance as $1/r^{1+\sigma}$,
defined by the Hamiltonian
\begin{equation}
H = - \sum_{i < j} \; \frac{J}{|i - j|^{1+\sigma}} \; \delta (s_i, s_j) \; , 
\label{eq:hamilt} 
\end{equation}
where  $J > 0$, $s_i$ denotes the $q$-state Potts variable at site $i$, $\delta$ is Kronecker 
symbol, and summation is taken over all pairs in the system.
The phase transition at nonzero temperature, shown rigorously \cite{DY} for the Ising $(q = 2)$ case 
with $\sigma \le 1$ and by renormalization group for the continuous n-component models with $\sigma < 1$ 
\cite{KO}, exists also in model (1) for $\sigma \le 1$ and $q>0$ \cite{CA,GU} and goes through a 
variety of universality classes by variation of $q$ and $\sigma$. Model (1) has been used as a 
relevant model for describing a number of
phenomena involving LR interactions, from spin glasses to neural networks,
but may also be of interest for possible analogies \cite{TS},
in some cases very direct \cite{LB96},
with short-range (SR) models in higher dimensions.

An important feature of Potts models with SR interactions is the
onset of the first-order phase transition for $q$ above
some threshold value $q_c(d)$, which depends on dimensionality \cite{WU}.
For example, in $d=2$ \cite{BA} and $d=4$ \cite{AP} analytical results yield $q_c$ equal to 
$4$ and $2$ respectively,
while for $d=3$, the approximate methods give a non-integer value for $q_c$,
slightly lower than $3$ \cite{BS}, so that the $3d$ $3$-state Potts model has an extremely
weak first-order transition, very difficult to detect \cite{JV}.
Such a distinct situations created by variation of $d$ and $q$ make this model
a canonical example for study of various aspects of temperature-driven first-order
phase transitions \cite{BI}.  

One can expect similar behaviour to occur for the Potts model with LR interactions,
although certain related quantities (e.g. the interface energy) may require different
interpretation.
Generally, this model has been much less explored than its SR version, due to
 non-locality of interactions, which makes difficult the application of
standard renormalization group techniques in direct space, 
otherwise appropriate for discrete models.
A few studies that have been done concern mostly the mean-field region and
its vicinity by using an $\epsilon-$expansion within Ginzburg-Landau continuous formalism \cite{PL}, 
and the special case of $\sigma=1$ \cite{CA}.    
We have recently proposed the finite-range scaling (FRS) approach \cite{UG}   
suitable for study of the Hamiltonian (1), with both $\sigma$ and $q$  
arbitrary and continuous \cite{GU}.  
This approach  is, however, inherently insensitive to discern the first-order transitions
and remained inconclusive with this respect.
The problem was not resolved in other recent works \cite{LB95,CM} on the LR Potts model either.

 For this reason we present here the results based on simple numerical 
simulations performed with intention to examine the existence of a first-order 
transition in this model and get a qualitative estimate of its dependence 
on $q$ and $\sigma$.

It has been recently pointed out \cite{LK}, that the 
temperature-driven first-order transitions can be identified from finite-size 
scaling (FSS) analysis of maxima of the energy probability distribution
\begin{equation}
P_L(E) = \frac{1}{Z_L(K)} {\cal N}_L(E)\ e^{-K E} \; , 
\end{equation}
where $K=J/k_BT$, $Z_L(K)$ is the partition function, ${\cal N}_L(E)$ is the number of 
configurations with energy $E$, and index $L$ denotes the system size. 
Due to coexistence of phases, at the temperature of a first-order phase transition
$P_L(E)$ has two maxima, corresponding to the two wells in the free energy.
The barrier separating them, which represents the interface free energy 
is defined by 
\begin{equation}
\Delta F_L  = \ln\left. \frac{P_L(E_{min})}{P_L(E_{max})}\right\vert_{K_L} \; , 
\label{eq:DelF} 
\end{equation}
where $E_{min}$ and $E_{max}$ denote the energies corresponding to the 
minimum and one of the two maxima respectively, the finite-size temperature  $K_L$ being 
adjusted so as to make the two maxima equal. For a first-order phase transition 
$\Delta F_L$ should diverge with $L$.
In systems with SR interactions it scales like a surface, 
i.e. $\sim L^{d-1}$, while in the present case it is expected to scale rather like a volume, i.e.
$\sim L$. 

The calculations were performed on chains of size $100 \le L \le 400$ 
with periodic boundary conditions. We have used the simple Metropolis 
single-spin-flip algorithm with $1\times 10^6 - 3\times 10^6$ Monte Carlo (MC) sweeps per spin. 
The number of necessary runs for the precise localization of each of
the size-dependent critical temperatures $K_L$ has been
reduced by applying the Ferrenberg and Swendsen \cite{FS} histogram method.

We have considered integer values of $q \ge 2$ in the 
interval $0 < \sigma \le 1$. While for $q=2$ (Ising case),  
the simulations at $T_c$ show only a single maximum in $P_L(E)$ in the entire range of $\sigma$, 
for higher values of $q$, the two peaks emerge for $\sigma$ sufficiently low. 
They become more pronounced with 
increasing $q$ or decreasing $\sigma$. For illustration, in Figure 1 are shown the 
maxima corresponding to different values of $\sigma$, taken from three typical sets
of simulations with fixed $q = 3$ and $L = 400$. 

We report here the systematic results for two chosen values, $q=3$ and $q=5$,
in the whole interval $0 < \sigma \le 1$ taken with increment $0.1$.

In Figures 2a and b are summarized the results for the free energy 
barrier plotted as a function of chain size $L$ for $q=3$ and $q=5$    
respectively. 
The corresponding critical temperatures extrapolated to $L\to\infty$, given in table 1, are 
found to be in good agreement with our earlier FRS results \cite{GU}, as well as with other 
known approximate results \cite{CM}. 

\begin{table}[hbt] 
\begin{center} 
\begin{tabular}{cccccccl} \hline \hline
 \\
 $\sigma$  & $q$ & $K_e^{MC}$ & $K_e^{FRS}$ & & $q$ & $K_e^{MC}$ & $K_e^{FRS}$
 \\
 \\ \hline \hline
 \\
 0.1 &3 &  0.190  & 0.136 & &  5 & 0.262      &  0.28  \\
 0.2 &  &  0.279  & 0.270 & &    & 0.333      &  0.45  \\
 0.3 &  &  0.380  & 0.386 & &    & 0.492      &  0.576  \\
 0.4 &  &  0.489  & 0.494 & &    & 0.637      &  0.690 \\
 0.5 &  &         &       & &    & 0.771      &  0.803   \\
 0.6 &  &         &       & &    & 0.901      &  0.920  \\
 0.7 &  &         &       & &    & 1.019      &  1.046     \\
 \\
\hline \hline
\end{tabular}
\end{center} 
\caption{ 
Inverse critical temperatures ($K_e^{MC}$), obtained by extrapolation of  $K_L$
compared to FRS extrapolated values ($K_e^{FRS}$) \protect\cite{GU}.
}
\end{table}

For both considered values of $q$, 
there is a wide range of $\sigma$, where $\Delta F_L$ increases with 
size, indicating the first-order transition.  
As expected for the LR interactions, $\Delta F_L$ is 
proportional to volume rather than surface and depends linearly on $L$. 
The slope is larger by an order 
of magnitude for $q=5$ in comparison to $q=3$, showing that the first-order 
character becomes stronger with increasing $q$, like for the SR interactions.
In both cases the slope decreases with increasing $\sigma$ to the point where 
$\Delta F_L$ becomes of the order of error bars. Beyond this value, at least 
for sizes considered here, the $P_L(E)$ exhibits a single maximum indicating  
the onset of the second-order phase transition. In present calculation, 
taken with a rough increment of 0.1 in $\sigma$, this change is observed 
around $\sigma=0.5$ and $\sigma=0.8$ for $q=3$ and $q=5$, respectively. 
These values should be taken with caution and only as a lower limit for the 
threshold value $\sigma_c$ between the first- and the second-order transition. Namely, the
present model allows the continuous approach to the threshold value $\sigma_c$,
whereby the first-order transition becomes arbitrarily weak and very difficult to detect, 
comparable to the situation with the $3d$ $3$-state Potts model with SR interactions. 
The above results, however, strongly suggest that $\sigma_c$ is considerably larger for $q=5$ 
than for $q=3$, and that dependence $\sigma_c(q)$ should be expected, analogous 
to threshold dependence $d_c(q)$ in the SR model.

Two other energy-related quantities are more conventionally \cite{BH} 
used for determination of the first-order 
transition in context of FSS analysis of MC simulation results: specific
heat and Binder's fourth order cumulant \cite{BIN}, which both can be derived from $P_L(E)$, 
and expressed in terms of  higher energy momenta
$ \langle E^n \rangle_L = \sum_E  E^n P_L(E) $ . 
Specific heat is given by
\begin{equation}
C_L =\frac{K^2}{L^d} \left( \; \langle E^2\rangle_L - \langle E\rangle_L^2 \; \right) \; .
\end{equation}
According to the FSS theory, for second-order transitions its maximum scales as
$C_L^{max}\sim L^{\alpha/\nu}$, 
where $\alpha$ and $\nu$ are the critical exponents 
of the specific heat and the correlation length respectively.
When the transition is of the first order, it scales as a volume, i.e. $C_L^{max} \sim L^d$. 
Instead of the Binder fourth cumulant $V_L^{(4)}=1-U_L^{(4)}/3$, we consider
here the ratio 
\begin{equation}
U_L^{(4)}  = \; \frac{ < E^4 >_L  }{ < E^2 >_L^2 }  \; . 
\end{equation}
For the first-order transitions, $\lim_{L \to\infty} U_L^{(4)} = 1$ when $T \ne T_c$,  
while at $T=T_c$ $\lim_{L \to\infty} U_L^{(4)}=const > 1$. 
For second-order transitions it always tends to one.

We present the behaviour of
those two quantities on two examples: $q=5,~\sigma=0.2$ and $q=3,~\sigma=0.8$, 
representative of the first- and second-order regimes, respectively. 

In Figure 3a and b one can observe two kinds of behaviour 
of $C_L^{max}$ in the 
two cases: linear and power law. The fit to the form 
$C_L^{max} \sim L^x$ in the latter case gives the value $x=0.24$ for $q=3,  \sigma=0.8$. 
The bare extrapolation error bars for $x$ are estimated to be of order of $10\%$. 
The hyper-scaling relation with substitution of FRS result \cite{GU} $\nu=1.74$ gives 
$\alpha/\nu = 0.15$. The difference can be attributed to the general difficulty 
in extracting the critical exponents from the specific heat, and 
to the fact that the calculated exponent is small and additional 
correction terms due to finite-size gain importance.  

The convergence of the maxima of $U_L^{(4)}$ with size is presented 
in figure 2((c) and (d)). The points for the Ising model with $\sigma=0.5$ are added as a 
reference for second-order transition behaviour. The case $q=3, \sigma=0.8$ 
shows, within numerical error bars, clear convergence towards $1$. The points for $q=5, 
\sigma=0.2$ converge towards a much larger value, which is approximately $2.4$
when we take into account larger values of $L$ and use the linear extrapolation. 

Thus, the two quantities confirm earlier conclusions based upon the behaviour of 
$\Delta F_L$. However, at present stage, we could not extract from these quantities any 
better precision in determination of  $\sigma_c(q)$, so we do not reproduce any 
systematic study for them.

In summary, by simple numerical calculations, in combination with FSS arguments,
we have shown that the $1d$ LR Potts 
model for integer $q > 2$ exhibits the first-order phase transition for 
$\sigma$ below some threshold value  $\sigma_c$ generally depending 
on $q$. First-order character becomes weaker with the increase of  $\sigma$, 
which represents a continuous parameter leading from first- to second-order
phase transition regime. More intensive numerical approach \cite{GUp} should 
be needed in future in order to determine the threshold value $\sigma_c(q)$.


\newpage

Figure captions: 
\vskip .25in

Figure 1:

Maxima of $\ln P_L(E')$ for $q = 3, L = 400$ and $\sigma = 0.3, 0.4$ and $0.5$, taken at 
respective values of $K_L$. $E' = E / |E_0(\sigma)|$, where the $E_0(\sigma)$ stay for the 
zero-temperature energies.

\vskip .25in

Figure 2:

MC results of the free energy barrier $\Delta F_L$ for sizes $L=100$ to $400$ 
for: (a) $q=3, \sigma=0.1$ to $0.4$, (b) $q=5, \sigma=0.1$ to $0.7$. Notice
that the slope is by an order of magnitude larger in the case (b).
The size of the numerical error bars is comparable to or smaller than the size of the points.

\vskip .25in

Figure 3:

MC data for sizes $L=100$ to $400$:
 specific heat maxima are plotted for
(a) $q=3, \sigma=0.8$, (b) $q=5, \sigma=0.2$; ratio $U_L^{(4)}$ maxima are  plotted in function of
inverse size for: (c) $q=3, \sigma=0.8$, (d) $q=5, \sigma=0.2$ (full circles). The diamonds
correspond to $q=2, \sigma=0.5$ taken as a reference. Lines show the linear extrapolations.
Notice the common scale on x-axis, but different scales on y-axis. 
The size of the numerical error bars is smaller than the size of the points.

\newpage

\pagestyle{empty}
\parindent 0pt
\null\vskip-40pt
\vskip 5 cm
\centerline{\bf Fazni prijelaz prvog reda u $1d$ Pottsovom modelu s dugodose\v znim medjudjelovanjem}
\vskip .25in
\centerline{K Uzelac i Z Glumac}
\vskip .25in
{\it Institut za Fiziku, Bijeni\v cka 46, POB 304,  10000 Zagreb, Hrvatska}
\vskip 2.5 cm
\centerline{\bf Sa\v zetak}

\baselineskip 21pt

U jednodimenzionalnom Pottsovom modelu s $q$ stanja i s dugodose\v znim 
me\-dju\-dje\-lo\-va\-nji\-ma koja 
opadaju s udaljeno\v s\v cu kao $1/r^{1+\sigma}$, Monte Carlo simulacijama je promatran fazni 
prijelaz prvog reda za $0 < \sigma \le 1$ i cjelobrojne vrijednosti $q > 2$.  
Na temelju {\it scaling} analize slobodne energije medjuplohe, specifi\v cne topline i 
Binderovog kumulanta \v cetvrtog reda, dobivamo prijelaz prvoga reda za $\sigma$ manji 
od grani\v cne vrijednosti $\sigma_c(q)$.

\end{document}